\documentclass[12pt]{article}

\usepackage[linktocpage=true,plainpages=false]{hyperref}
\usepackage{color}
\usepackage{amsmath,amssymb}
\usepackage{cite}
\usepackage{array}
\usepackage[T1]{fontenc}
\usepackage[labelsep=period]{caption}

\definecolor{linkcolor}{rgb}{0.6,0,0}
\definecolor{citecolor}{rgb}{0,0.6,0}
\definecolor{urlcolor}{rgb}{0,0,0.9}
\hypersetup{colorlinks, linkcolor={linkcolor},citecolor={citecolor}, urlcolor={urlcolor}}

\newcommand{\dd}{\partial}
\newcommand{\de}{\delta}
\newcommand{\m}{\mu}
\newcommand{\n}{\nu}
\newcommand{\ls}{\left(}
\newcommand{\rs}{\right)}

\newcommand{\la}{\lambda}
\newcommand{\ka}{\varkappa}
\newcommand{\tr}[1]{\overset{{\scriptscriptstyle 3}}{#1}{}}
\newcommand{\pos}{\tr{\Pi}_{\!\!\bot}{}}
\newcommand{\pua}[2]{\left\{#1,#2\right\}}
\newcommand{\dz}{\zeta}

\newcommand{\ga}{\gamma}
\newcommand{\si}{\sigma}

\newcommand{\disn}[2]{$$\displaylines{\refstepcounter{equation}%
            \label{#1}\hskip 1em minus 1em #2\hfilneg}$$}
\newcommand{\nom}{\hfil\hskip 1em minus 1em (\theequation)}
\newcommand{\no}{\hfil \hskip 1em minus 1em\phantom{(\theequation)}%
            \hfilneg\cr\hfilneg\hskip 1em minus 1em\hfil}
\newcommand{\ns}{\hfill\cr\hfill}

\begin{document}
\title{Algebra of Implicitly Defined Constraints for Gravity\\
as the General Form of Embedding Theory}
\author{S.~A.~Paston\thanks{E-mail: s.paston@spbu.ru}, E.~N.~Semenova\thanks{E-mail: derenovacio@mail.ru}, V.~A.~Franke\thanks{E-mail: v.franke@spbu.ru}, A.~A.~Sheykin\thanks{E-mail: a.sheykin@spbu.ru}
\\
{\it Saint Petersburg State University, Saint Petersburg, Russia}
}
\date{\vskip 15mm}
\maketitle

\begin{abstract}
We consider the embedding theory, the approach to gravity proposed by Regge and Teitelboim,
in which 4D space-time is treated as a surface in high-dimensional flat ambient space. In its general form,
which does not contain artificially imposed constraints, this theory can be viewed as an extension of GR.
In the present paper we study the canonical description of the embedding theory in this general form. In
this case, one of the natural constraints cannot be written explicitly, in contrast to the case where additional
Einsteinian constraints are imposed. Nevertheless, it is possible to calculate all Poisson brackets with this
constraint. We prove that the algebra of four emerging constraints is closed, i.e., all of them are first-class
constraints. The explicit form of this algebra is also obtained.
\end{abstract}

\newpage

\section{Introduction}
In 1975, T. Regge and C. Teitelboim suggested an approach to the description of the theory of gravity \cite{regge}, similar in formulation to the description of a relativistic string. They supposed that our space-time
is a four-dimensional surface in flat ten-dimensional space. The independent variable describing gravity
is not a metric $g_{\m\n}(x^\ga)$, but the embedding function
	$y^a(x^\ga)$  of this surface to the ambient space (here and
henceforth $\m,\n,\ldots=0,1,2,3$, $a,b,\ldots=0,1,\ldots,9$).
The metric then becomes induced and expressed through the
embedding function:
\disn{v1}{
g_{\mu\nu}=(\partial_\mu y^a) (\partial_\nu y^b)\eta_{ab}
\nom}
(here $\eta_{ab}$ is the flat metric of the ambient space), therefore this approach to the description of gravity
is called the embedding theory.

The number of dimensions of the embedding space
(the bulk) is chosen in accordance with the Janet-Cartan-Friedman theorem \cite{fridman61}, which states that an
arbitrary pseudo-Riemannian space of dimension n
can be locally isometrically embedded into an arbitrary pseudo-Riemannian space (including flat space)
whose dimension is greater than or equal to
$N=\dfrac{n(n+1)}{2}$
and it contains a number of timelike and spacelike
directions not less than there is in the embedded
space.
It is convenient to assume that there is only one
timelike direction in the bulk, i.e., the bulk is ten-dimensional Minkowski space $R^{1,9}$ (in this paper, we
will use the signature $(+,-,-,\dots,-)$).

After formulation of the embedding theory in \cite{regge},
this approach was discussed, in particular, critically,
in \cite{deser}. Later on, the idea of embedding into a flat
space was repeatedly used in the description of gravity \cite{pavsic85let,tapia,maia89,bandos,estabrook1999,davkar,statja18,faddeev,statja25,willison}, and in some cases this idea emerged independently. A fundamental difference between the
embedding theory from rather similar approach of brane theory \cite{lrr-2010-5} is the lack of gravity in the bulk because it is postulated that it is flat. In the framework of the embedding ideas, there were also attempts to associate quantum effects in the spacetime and in the bulk \cite{deserlev99,banerjee10,statja34} (see also references in \cite{statja34}), for which purpose constructed were explicit
embeddings of physically interesting solutions of the Einstein equations, see, for example, \cite{collinson68,davidson,statja27,statja30} and
references in \cite{statja30}.

The description of gravity in the framework of the embedding theory may be more convenient for building a quantum theory of gravity since in this approach, in a natural way, there emerges flat space. Many of the problems that appear at attempts to
quantize gravity in terms of the metric (their presentation can be found, for example, in the review \cite{carlip})
arise from the fact that we are trying to apply the quantization procedure that has proven to be successful for field theory in flat space, to the case where the dynamic variables are geometric characteristics of this space, and it is these quantities that are subject
to quantization. A brief discussion of the ways of solution of some of the existing important problems in the approach of the embedding theory are described in the introduction of the the recent paper \cite{statja35}.

As the action of gravity in the embedding theory one usually takes the standard Einstein-Hilbert expression where the metric is regarded to be expressed in terms of the embedding function $y^a(x)$, which plays the role of an independent variable, by \eqref{v1}. Inclusion of arbitrary forms of matter is then implemented in the usual way, by adding the corresponding action
of matter,
\disn{v3}{
S=\int d^{4}x \sqrt{-g}\ls-\frac{1}{2\ka}\, R+{\cal L}_m\rs,
\nom}
where $\ka$ is the Einstein gravitational constant, $R$
is the scalar curvature, and ${\cal L}_m$ is the Lagrangian
density of matter. Despite the same form of the action
as in GR, the equations of motion in the embedding
theory turn out to be more general than the Einstein
equations. A reason for that is the existence of derivatives in Eq. (\ref{v1}) which realizes the transition to a new
independent variable.

The equations emerging from variation in $y^a(x)$
(usually called the Regge-Teitelboim equations) may
be written in one of the two equivalent forms: either
in the form making it evident that the equations do
not contain higher than second-order derivatives of
the independent variable $y^a(x)$:
\disn{v4}{
(G^{\mu\nu}-\varkappa\, T^{\mu\nu})b^a_{\m\n}=0
\nom}
(it can be shown that $G^{\mu\nu}$ are expressed through first
and second-order derivatives of $y^a(x)$), or in the form
of a kind of continuity equation
\disn{v5}{
\partial_\mu \Bigl(\sqrt{-g}(G^{\mu\nu}-\varkappa\, T^{\mu\nu})\partial_\nu y^a\Bigr)=0.
\nom}
Here $G^{\mu\nu}$ is the Einstein tensor, $T^{\mu\nu}$ -- is the energy-momentum tensor of matter,
$b^a_{\m\n}=D_\m \dd_\n y^a$ is the
second fundamental form of the surface, and  $D_\m$is a
covariant derivative, see details in  \cite{statja18}.
The equivalence of Eqs. (\ref{v4}) and (\ref{v4}) follows from the identity $D_\m G^{\m\n}=0$ and the covariant conservation law$D_\m T^{\m\n}=0$, which follows from the equations of motion of matter.

It is easy to notice that all solutions of the
Einsteins equations are solutions of the Regge-Teitelboim equations, but the reverse is not true.
The latter can contain so-called "extra" (that is,
non-Einsteinian) solutions, and consequently the
embedding theory is not equivalent to GR but is its
extension. One can try to use the extra solutions
to explain some effects which in the framework of
GR require introduction of specific types of matter,
such as dark energy and dark matter, see \cite{davids97,davids01,statja26}. However, considering the presence of the inflationary
era in the history of the Universe, there is a reason to
believe that, at least under the Friedmann symmetry,
the extra solutions are very strongly suppressed \cite{statja26}.

It is possible not to treat the embedding theory as
an extension of GR but, instead, to try to find such
a form of the embedding theory in which it would be
equivalent to GR. It is this task that was posed in the
original paper \cite{regge}. To solve it, it was suggested to
impose, in addition, the Einstein constraints, four of
the ten Einstein equations
\disn{v6}{
G^{\mu 0}-\ka\, T^{\mu 0}=0.
\nom}
It is sufficient to impose these constraints only at the
initial time instant \cite{statja18}, and, as a result, the "full" embedding theory (containing extra solutions), is re-
duced to a more narrow theory, equivalent to GR. It
may be of interest to study both versions of the theory,
the one equivalent to GR (one can call it "the Regge-Teitelboim formulation of GR", or RT-GR) and the
full version, containing extra solutions and being an
extension of GR.

Of greatest interest are the advantages that may
arise at quantization in the description of gravity in
the form of the embedding theory due to the natural
appearance of flat space in this approach. One of
the important steps on the way to quantization of
the theory is the construction of its canonical (i.e.,
Hamiltonian) description. In this case, since in the
canonical formalism for gravity there always emerge
constraints (for GR they are usually studied in the
Arnowitt-Deser-Misner (ADM) approach \cite{adm}), it is
significant whether these constraints belong to the
first class by Dirac's classification, i.e., whether they
form an algebra of first-class constraints. The present
paper is devoted to answering this question for the
"full" embedding theory.

In Section 2 we construct the canonical formalism
of the theory and discuss the difficulties emerging in
its construction. A basic one is the inability to write
down one of the constraints explicitly in the "full"
embedding theory. Note that in \cite{davkar} it was suggested to
introduce an auxiliary field, allowing one to overcome
this difficulty, however, as a result, there emerged
second-class constraints in the theory, which requires
using the complex formalism of Dirac brackets. In
Section 2 we also demonstrate the differences in the
canonical formalism of the "full" embedding theory
from the previously studied case of additionally imposing the Einstein constraints (\ref{v6}) in the RT-GR
formulation. Section 3 presents a calculation of Poisson brackets of all constraints with each other, and
it is proved that the constraints belong to the first
class, and an explicit form of the emerging algebra of
constraints is found. This becomes possible despite
the impossibility to write down one of the constraints
explicitly. This result can be used at attempts to
quantize gravity in the form of the "full" embedding theory by writing a functional integral in the canonical variables.
\section{Canonical formalism
of the "full" embedding theory}\label{kan}
In the construction of the canonical formalism,
time turns out to be distinguished (the so-called 3+1
splitting takes place), therefore the 4D space-time
turns out to be presented as a system of 3D surfaces
of constant time. These surfaces are automatically
embedded into the bulk  $R^{1,9}$
and are described by the
embedding functions
\disn{1.6}{
y^a(x^i)\equiv y^a(x^\m)|_{x^0=const},
\nom}
here and henceforth $i,k,\ldots=1,2,3$. For such 3D
surfaces it is useful to introduce the quantities that
characterize their geometry. We will mark them by
the symbol "3" to distinguish them from the corresponding quantities for the 4D surface corresponding
to the whole space-time.
Let us introduce the "non-quadratic vierbein"
\disn{1.6.1}{
\tr{e}^a_i=\dd_i y^a,
\nom}
in terms of which the 3D metric is expressed by the
relation
\disn{1.6.1a}{
\tr{g}_{ik}=\tr{e}^a_i \tr{e}^b_k\eta_{ab},
\nom}
as well as the "inverse non-quadratic vierbein"
 $\tr{e}_a^i$ and
the projectors:
$\tr\Pi^a_b$ onto the tangent subspace at a
given point, and
$\pos^a_b$ onto its orthogonal subspace:
\disn{1.6.2}{
\tr{e}_a^i=\tr{e}^b_k \tr g^{ki} \eta_{ba},\qquad
\tr\Pi^a_b=\tr{e}^a_i \tr{e}_b^i,\qquad
\pos^a_b=\de^a_b-\tr\Pi^a_b.
\nom}
We will also use, in what follows, the second fundamental form of the 3D surface in question:
\disn{1.6.3}{
\tr b^a_{ik}=\tr{D}_i\dd_k y^a=
\pos^a_b\dd_i\dd_k y^b.
\nom}
The formalism of the embedding theory is given in
more detail in \cite{statja18}, and one can find there all necessary explanations to the relations presented here.

In the study of the canonical formalism of the
embedding theory with the action (\ref{v3}), first of all
there emerges a problem connected with the fact that
the corresponding Lagrangian contains second-order
time derivatives of the independent variable $y^a(x)$ \cite{tapia}.

Therefore, to construct a canonical formalism, it
is necessary to invoke special methods. In \cite{tapia},
for this purpose, the Ferraris-Francavilla method
was used \cite{ferraris_canon}, and later in \cite{rojas04,rojas06,rojas09,rojas13} the Hamilton-Ostrogradsky approach (taking into account the surface term) was applied to the embedding theory. However, if for the action of gravity, instead of the
Einstein-Hilbert one, we use the ADM action \cite{adm}
which differs from it by the surface term only, then the
second-order time derivatives of $y^a(x)$ disappear from
the action \cite{regge,frtap}. As a result, the action acquires the
form (for brevity, we omit the contribution of matter
and the common factor $-1/(2\ka)$  which is insignificant for the canonical analysis of the theory)
\disn{k1}{
S=\int d^{4}x \sqrt{-g} \ls\tr{R} +(K^i_i)^2-K_{ik}K^{ik}\rs,
\nom}
where $\tr{R}$ is the scalar curvature of the 3D surface $x^0=const$, and $K_{ik}$  is the second quadratic form
of this surface in 4D space-time, and the indices of
this quantity are raised and lowered by the 3D metric
 (\ref{1.6.1a}).
With this choice of the action, the canonical
formalism may be constructed in the usual way.

The action (\ref{k1}) may be rewritten in the form in
which all occurences of the generalized velocity $y^a$
(the dot means a derivative with respect to time $x^0$).
This form reads (a derivation of this relation is presented in\cite{statja18}):
\disn{2.1}{
S=\int dx^0 L(y^a,\dot y^a),
\qquad
L=\frac{1}{2}\int d^3x\ls
\frac{\dot y^a B_{ab}\;\dot y^b}{\sqrt{\dot y^c\pos_{cd}\,\dot y^d}}+
\sqrt{\dot y^c\pos_{cd}\,\dot y^d}\;B^a_a\rs,
\nom}
where
\disn{2.2}{
B^{ab}=2\sqrt{-\tr g}\;\tr b^a_{ik}\tr b^b_{lm}\ls \tr g^{ik}\tr g^{lm}-\tr g^{il}\tr g^{km}\rs.
\nom}
From  (\ref{2.1})  one can easily find the generalized momentum for the independent variable $y^a$
\disn{2.3}{
\pi_a=\frac{\de L}{\de \dot y^a}=
B_{ab}n^b-\frac{1}{2}n_a\ls n_c B^{cd} n_d-B^c_c\rs,
\nom}
where
\disn{2.4}{
n^a=\frac{\pos^a_b\,\dot y^b}{\sqrt{\dot y^c\pos_{cd}\,\dot y^d}}
\nom}
is the unit vector tangent to the 4D space-time surface at a given point, which is orthogonal to the 3D
surface $t = const$.

From the definition (\ref{2.3}) of the generalized momentum it follows that there are some primary constraints. Using Eq. (\ref{2.2}) as well as the transversality
properties of the quantities $\tr b^a_{ik}$ and $n^a$ by the index $a$ that follows from (\ref{1.6.3}) and  (\ref{2.4}), it is easy to obtain
three primary constraints:
\disn{3.1}{
\Phi_i=\pi_a\tr{e}^a_i\approx 0.
\nom}
(here and henceforth the symbol $\approx$ means a "weak"
equality which cannot be used until all Poisson
brackets have been calculated). In the full theory, in
which the Einstein constraints (\ref{v6}) are not additionally
imposed, one more constraint emerges as a consequence of the vector $n_a$ being normalized to unity:
\disn{3.2}{
 \Phi_4=n^a(\pi,y)n_a(\pi,y)-1\approx 0,
\nom}
where  $n^a(\pi,y)$ denotes a solution of  (\ref{2.3}) with re
spect to $n^a$ (in general, also defined at the values of the momentum $\pi^a$ which do not satisfy the constraints(\ref{3.1})).

However, for RT-GR, where the Einstein constraints are imposed, the form of the fourth constraint
substantially changes \cite{statja18} (although in \cite{regge} it was by error also written in the form (\ref{3.2})). The point is that if the Einstein constraints hold, the expression for the
generalized momentum takes the form
\disn{2.3.1}{
\pi_a=B_{ab}n^b,
\nom}
while the matrix $B_{ab}$ has a rank not higher than 6, and
therefore the normalization of the vector $n^a$ no more
imposes a restriction on the generalized momentum
$\pi_a$, see details in \cite{statja18}. As a whole, in the canonical formalism of RT-GR there are eight constraints:
four primary ones and four additionally imposed Einsteinian ones \cite{statja18}. In \cite{statja24} all Poisson brackets of
the constraints with each other were calculated, and
an algebra they form was found. All of them turned
out to be first-class constraints because the results
of Poisson brackets calculations are proportional to
the constraints. Thus, even though some of the
constraints were imposed by hand, they turned out
to be in involution with the dynamics of the theory.
In \cite{statja35} it has been verified that this property is also
preserved when one partially fixes the gauge in the
action by identifying the internal and external times, $y^0(x^\m)=x^0$.

Let us return to the "full" embedding theory where
the Einstein constraints are not imposed and the
fourth primary constraint has the form  (\ref{3.2}). In this
case it cannot be written as an explicit function of the
canonical variables since (\ref{2.3}), being a multidimensional cubic equation with respect to $n^a$, has no
explicit solution.
It is easy to verify that if the primary
constraints  (\ref{3.1}) and (\ref{3.2}) hold,  the Hamiltonian of the theory turns out to be equal to zero,
\disn{3.4}{
H=\int d^3x\;\pi_a\dot y^a-L \approx 0,
\nom}
hence secondary constraints do not emerge. As a result, the generalized Hamiltonian reduces to a linear
combination of the constraints $\Phi_i$ and $\Phi_4$,
and the latter contains the function  $n^a(\pi,y)$ specified implicitly.

\section{The constraints algebra}\label{alg}
Let us calculate the Poisson brackets of the existing four constraints (\ref{3.1}), (\ref{3.2})  with each other. We will show that, despite the implicit form of the constraint (\ref{3.2}), it is still possible to calculate the Poisson brackets of this constraint with other quantities. To do that, we note that when calculating the Poisson
brackets, we must find the derivatives of the implicit
functions $\de n(\pi,y)/\de \pi$ and $\de n(\pi,y)/\de y$,
and it can be done using the formula for differentiating an implicit function
\disn{3.2.1}{
\frac{\de n}{\de\pi}=\ls\frac{\de\pi}{\de n}\rs^{-1}.
\nom}
Let us implement this idea.

For convenience of the calculations, let us introduce for all constraints their contractions with arbitrary finite functions:
\disn{3.5}{
\Phi_\xi=\int d^3x\,\Phi_i\,\xi^i,\qquad
\Phi_\ga^4=\int d^3x \,\Phi_4\,\ga.
\nom}
Note that relative to three-dimensional coordinate
transformations the generalized momentum is a
scalar density, therefore, $\Phi_i$ is a vector density, while $\Phi_4$ is a scalar.  Therefore, we should assume that the
arbitrary function $\xi^i$ is a vector while $\ga$ is a scalar
density.
It is known \cite{statja18} that the constraint $\Phi_\xi$ is a generator of 3D transformations since
 \disn{3.6}{\pua{
\Phi_\xi}{y^a}=\xi^i\dd_iy^a,\qquad
\pua{\Phi_\xi}{\frac{\pi_a}{\sqrt{-\tr{g}}}}=\xi^i\dd_i\frac{\pi_a}{\sqrt{-\tr{g}}}.
\nom}
It is therefore not difficult to find all Poisson brackets
that contain $\Phi_\xi$: the action of this constraint on any
quantity is uniquely determined by the transformation
properties of this quantity at 3D coordinate transformations. The result is
\disn{3.7}{
\pua{\Phi_\xi}{\Phi_\dz}=-\int d^3x\,\Phi_k\ls\xi^i\dd_i\dz^k-\dz^i\dd_i\xi^k\rs,\qquad
\pua{\Phi_\xi}{\Phi_\ga^4}=\int d^3x\,\ga\,\xi^k\dd_k\Phi_4.
\nom}

The calculation of the remaining Poisson brackets $\pua{\Phi^4_\ga}{\Phi^4_\si}$ is a nontrivial task, and here it is necessary
to use the above-described idea. The Poisson bracket
of two constraints  (\ref{3.2}) has the form
\disn{3.3}{
\pua{n_a(x)n^a(x)-1}{n_b(\tilde x)n^b(\tilde x)-1}=
4\int d^3 \hat x \ls n_a(x)\frac{\de n^a(x)}{\de \pi_c(\hat x)}n_b(\tilde x)\frac{\de n^b(\tilde x)}{\de y^c(\hat x)}-
(x\;\;\leftrightarrow \;\;\tilde x)\rs,
\nom}
where $(x\;\leftrightarrow \;\tilde x)$ denotes the same expression with interchanged  $x$ and $\tilde x$. To calculate this expression, it is necessary to learn how to find the involved variational derivatives. To do that, let us first of all rewrite Eq. (\ref{2.3}) in the form
\disn{3.8}{
\pi_a=\frac{1}{2} C_{abc}B^{bc},
\nom}
where the tensor
\disn{3.9}{
C_{abc}=\ls n_a\eta_{bc}+ n_b\eta_{ac}+ n_c\eta_{ab}- n_a n_b n_c\rs
\nom}
is completely symmetric. Using (\ref{3.8}), let us write
down the variation of the generalized momentum $\pi^a$
at arbitrary variations of the embedding function $y^a$
and the vector $n^a$ (we stress that the variation of $n^a$
is quite arbitrary, i.e., it may imply that the constraint
(\ref{3.1}) will no longer hold):

\disn{3.10}{
\delta\pi^a(x)=\frac{1}{2}C^{abc}(x)\int\! d \tilde x\, \frac{\de B_{bc}(x)}{\de y^d( \tilde x)}\delta y^d( \tilde x)+A^a_{\;\;b}(x)\delta n^b(x),
\nom}
where we have used the notations
\disn{3.11}{
A^a_{\;\;b}=-\chi\de^a_b+(\de^a_c-n^an_c)B^c_{\;\;b},\qquad
\chi\equiv\frac{1}{2}\ls n_cB^{cd}n_d-B^c_c\rs.
\nom}

It follows from Eq. (\ref{3.10}) that the variational derivative of the vector $n^b$ in $\pi^a$ at fixed $y^a$ has the form
\disn{3.12}{
\frac{\de n^b(x)}{\de\pi^a(\hat x)}=A^{-1b}_{\;\;\;\;\;\;a}(x)\de(x-\hat x),
\nom}
where $A^{-1}$ is the inverse matrix, while the variational
derivative of the vector $n^b$ in $y^a$ at fixed $\pi^a$
 has the form
\disn{3.13}{
\frac{\de n^b(x)}{\de y^a(\hat x)}=-\frac{1}{2}A^{-1b}_{\;\;\;\;\;\;c}(x)C^{cde}(x)
\frac{\de B_{de}(x)}{\de y^a(\hat x)}.
\nom}
Note that the matrix $A^a_{\;\;b}$ is invertible if $\chi\ne0$. We
assume that we are considering the generic situation where it is the case. This is the distinction of
the present version of the canonical formalism of the
"full" embedding theory from that studied in~\cite{regge,statja18,statja24}
where the Einstein constraints were additionally
imposed, since the quantity $\chi$ exactly coincides with
one of them, see \cite{statja18}.

As is evident from Eqs. (\ref{3.12}), (\ref{3.13}), to calculate
the expression (\ref{3.3}) we should learn how to find the
quantity $n_b A^{-1b}_{\;\;\;\;\;\;a}$.
To do this, we notice that
\disn{3.13.1}{
n_a A^a_{\;\;b}=-\chi n_b+ n_c B^c_{\;\;b}(1-n_d n^d),
\nom}
(Eq. (\ref{3.11}) has been used), whence
\disn{3.13.2}{
n_b A^{-1b}_{\;\;\;\;\;\;a}=-\frac{1}{\chi}\ls n_a+n_c B^c_{\;\;b}A^{-1b}_{\;\;\;\;\;\;a}\Phi_4\rs
\approx -\frac{1}{\chi} n_a,
\nom}
where we have taken into account the definition of
the constraint (\ref{3.2}). Thus we can see that the quantity $A^{-1b}_{\;\;\;\;\;\;a}$, which cannot be obtained explicitly, will
not enter into the result of calculating the Poisson
bracket (\ref{3.3})
on the constraint surface (that is, if the
constraints hold), and this substantially simplifies the
analysis of their closure.

Using (\ref{3.12}), (\ref{3.13}) and (\ref{3.13.2}), we can write down the expression (\ref{3.3}) as follows:
\disn{3.13.3}{
-\frac{2}{\chi(x)}\Bigl( n^c(x)+n_a(x) B^a_{\;\;b}A^{-1bc}(x)\Phi_4(x)\Bigr)\times\ns\times
\frac{1}{\chi(\tilde x)}\Bigl( n_f(\tilde x)+n_g(\tilde x) B^g_{\;\;h}(\tilde x)
A^{-1h}_{\;\;\;\;\;\;f}(\tilde x)\Phi_4(\tilde x)\Bigr)
C^{fde}(\tilde x)\frac{\de B_{de}(\tilde x)}{\de y^c(x)}-
(x\;\;\leftrightarrow \;\;\tilde x)\approx\no\approx
-\frac{2}{\chi(x)\chi(\tilde x)} n^c(x)n_f(\tilde x)
C^{fde}(\tilde x)\frac{\de B_{de}(\tilde x)}{\de y^c(x)}-
(x\;\;\leftrightarrow \;\;\tilde x).
\nom}
A further analysis of this result on the constraints
surface does not require dealing with implicitly specified expressions and can be performed by a direct
calculation. As a result of this analysis, it becomes
possible to prove that if the constraints  (\ref{3.1}) and (\ref{3.2})
the expression (\ref{3.13.3}), and hence the expression
 (\ref{3.3})
turn to zero, i.e., the Poisson brackets $\pua{\Phi^4_\ga}{\Phi^4_\si}$ turn out to be proportional to the constraints.

To obtain the exact form of the result of calculation
of these Poisson brackets, it is helpful to prove the
following property of the quantity $A^{-1b}_{\;\;\;\;\;\;a}$.
Note that from the definition (\ref{3.11})  it can be concluded that
\disn{3.15}{
A^c_{\;\;b}\tr{\Pi}^b_d=\ls-\chi\de^c_b+(\de^c_f-n^cn_f)B^f_{\;\;b}\rs\tr{\Pi}^b_d=-\chi\tr{\Pi}^c_d,
\nom}
where we have used the transversality condition following from (\ref{2.2}) and (\ref{1.6.3})
\disn{3.15.1}{
B^c_{\;\;f}\tr{\Pi}^f_b=0.
\nom}
Multiplying the relation (\ref{3.15}) by $A^{-1b}_{\;\;\;\;\;\;c}$, we find that
\disn{3.17}{
\tr{\Pi}^b_d=-\chi A^{-1b}_{\;\;\;\;\;\;c}\tr{\Pi}^c_d\qquad\Longrightarrow\qquad
A^{-1b}_{\;\;\;\;\;\;c}\tr{\Pi}^c_d=-\frac{1}{\chi}\tr{\Pi}^b_d
\nom}
under the assumption $\chi\ne0$.
It is also useful to note
that multiplying the equality  (\ref{2.3}) by $\tr{e}^a_i$ and using (\ref{3.15.1})
and the definitions (\ref{3.11}), (\ref{3.1})
 it is easy to obtain under the same assumptions that
\disn{3.19}{
\tr{e}^a_in_a=-\frac{1}{\chi}\pi_a\tr{e}^a_i=-\frac{1}{\chi}\Phi_i.
\nom}

Using these relations, as a result of cumbersome
calculations, it is possible to obtain the exact form of
the Poisson brackets:
\disn{9.1}{
\pua{\Phi_\si^4}{\Phi_\ga^4}=8\int d^3x\;\sqrt{-\tr{g}}\Bigg({\bf \Phi_j}\bigg[\frac{1}{\chi}\tr{e}^j_a\dd_k(\si Z^a)\dd_i(\ga Z^b) Y^{ik,c}(n_bn_c-\eta_{bc})+\ns
+\sqrt{-\tr{g}}\ga\dd_k\genfrac{(}{)}{}{1}{\si}{\sqrt{-\tr{g}}} \Big(-Y^{ik,b}\ls\frac{n^an_a}{\chi^2}\tr{b}_{bi}^{\;\;\;j}+\frac{2}{\chi^2}\tr{b}^{a\;\;j}_{\;\;i}n_an_b(2-n^cn_c)\rs\Big)+\ns
+\frac{1}{\chi}\ls-\tr{g}^{ik}Z^a\ls n_a+\chi Z_a\rs-\frac{2}{\chi}\sqrt{-\tr{g}}\tr{b}^a_{ip}\tr{b}^b_{lm}C_{cab}Z^c
\tr{g}^{jm}\ls\tr{g}^{ip}\tr{g}^{lk}+\tr{g}^{il}\tr{g}^{pk}\rs\rs\bigg]+\no
+{\bf \Phi_4}\frac{1}{\chi}\sqrt{-\tr{g}}\ga\dd_k\genfrac{(}{)}{}{1}{\si}{\sqrt{-\tr{g}}} Y^{ik,b}X^e\Big[ Z^a\dd_i\ls C_{edf}\pos^d_a\rs
-\frac{1}{\chi}\tr{e}^d_j\tr{b}_{ei}^{\;\;\;j}(n^an_a(\eta_{db}-n_dn_b)+2n_dn_b)\Big]-\ns
-\frac{1}{2\chi}\sqrt{-\tr{g}}\ga\dd_k\genfrac{(}{)}{}{1}{\si}{\sqrt{-\tr{g}}} Y^{ik,b}Z^a\pos^c_a(\eta_{cb}-n_cn_b)\dd_i{\bf \Phi_4}
\Bigg)-(\si\leftrightarrow\ga).
\nom}
Here, for compactness of the presentation, we have
used the following notations:
\disn{1.1}{
X^e=n_cB^c_fA^{-1f\;e},\qquad
Y^{ik,a}=\tr{b}^a_{lm}\ls\tr{g}^{ik}\tr{g}^{lm}-\tr{g}^{il}\tr{g}^{km}\rs,\no
Z_a=n_bA^{-1b}_{\;\;\;\;\;\;a}=-\frac{1}{\chi}\ls n_a+n_c B^c_{\;\;b}A^{-1b}_{\;\;\;\;\;\;a}\Phi_4\rs.
\nom}
As is easy to notice, each term in the right-hand side
of Eq. (\ref{9.1}) is proportional to one of the constraints.

Thus it has been possible to calculate the constraints algebra of the "full" embedding theory (without imposing the Einsteinian constraints) and to
prove that the algebra of the constraints $\Phi_i$, $\Phi_4$ is
closed, which means that these constraints are first-class. As a result, the generalized Hamiltonian of the "full" embedding theory reduces to a linear combination of the first-class constraints with the Lagrange multipliers:
\disn{1.1.1}{
H^{\text{gen}}=\int d^3x \ls \la^i \Phi_i+\la^4 \Phi_4\rs=
\int d^3x \ls \la^i \tr{e}^a_i \pi_a+\la^4 \ls n^a(\pi,y)n_a(\pi,y)-1 \rs\rs.
\nom}
When using this Hamiltonian, there is an inconvenience related to the implicit form of its dependence on the canonical variables $y^a$, $\pi_a$ (recall
that the function  $n^a(\pi,y)$ was defined as a solution
of the equation (\ref{2.3}) which could not be written in an
explicit form). Nevertheless, one can try to use the
Hamiltonian obtained for writing the corresponding
functional integral in the canonical variables with the
purpose to quantize the theory, for example, employing the opportunity to pass on from integration in the
generalized momentum $\pi_a$ to integration in $n^a$; in the
spirit of the idea recently put forward in \cite{cahill}.

{\bf
Acknowledgements.}
The work was supported by SPbU grant N 11.38.223.2015.


\end{document}